\documentclass[prd,aps,showpacs,nofootinbib,superscriptaddress,floatfix]{revtex4}
\usepackage{epsfig}
\usepackage{amsmath, slashed,graphicx}

\newcommand{\sT}{{\scriptscriptstyle T}}

\def\slash#1{\setbox0=\hbox{$#1$}  
   \dimen0=\wd0     
   \setbox1=\hbox{/} \dimen1=\wd1  
   \ifdim\dimen0>\dimen1   
      \rlap{\hbox to \dimen0{\hfil/\hfil}} 
      #1     
   \else     
      \rlap{\hbox to \dimen1{\hfil$#1$\hfil}} 
      /      
   \fi}      %

\begin{document}

\title{ Transverse Force Tomography}

\author{Fatma P. Aslan, Matthias Burkardt, Marc Schlegel }
 \affiliation{Department of Physics, New Mexico State University, Las Cruces, NM 88003-0001, U.S.A.}

\date{\today}

\begin{abstract}
While twist-2 GPDs allow for a determination of the distribution of partons on the transverse plane, twist-3 GPDs contain quark-gluon correlations that provide information about the average transverse color Lorentz force acting on quarks. We demonstrate how twist-3 GPDs can be used to provide transverse position information about that force.
\end{abstract}

\maketitle

\section{Introduction} 
\label{s:intro}
While twist-2 parton distribution functions (PDFs) provide information about the longitudinal momentum distribution of partons,  two-dimensional Fourier transforms of twist-2 generalized parton distribution functions (GPDs) for a vanishing skewness parameter $\xi=0$ provide information on the longitudinal momentum distribution of partons in the transverse plane (impact parameter space) \cite{Burkardt:2000za}, i.e.
\begin{equation}\label{FTtwist2}
q(x, {\bf b}_{\perp})=\int{}\dfrac{d^2 {\bf \Delta}_{\perp}}{(2\pi)^2}e ^{-i{\bf b}_{\perp}\cdot {\bf \Delta}_{\perp}}\hspace{.1 cm} \mathrm{GPD^{twt-2}}(x,-{\bf \Delta}_{\perp}^2),
\end{equation}
where $q(x, {\bf b}_{\perp})$ is the impact parameter parton distribution as a function of the separation (${\bf b}_{\perp}$)  from the transverse center of momentum 
$({\bf R}_{\perp}\equiv\sum_{i=q,g}{\bf r}_{\perp,i}x_i)$. 

On the other hand, twist-3 distributions involve quark-gluon correlations that are not contained in  twist-2 distributions. Even though they do not have a single particle density interpretation like twist-2 distributions, it has been shown that the $x^2$ moments of \textit {intrinsic} twist-3 PDFs are related to the quark-gluon correlations which have a further interpretation as a \textit{force} \cite{Burkardt:2008ps}. For example, the chirally-even spin-dependent twist-3 parton distribution $g_\sT^q(x)$, defined as (see, e.g., Ref. \cite{Kanazawa:2015ajw})
\begin{eqnarray}
-M\,g_\sT^q(x)=\int_{-\infty}^\infty\frac{\mathrm{d}\lambda}{4\pi}\,\mathrm{e}^{i\lambda x}\langle P,S_\sT|\,\bar{q}(0)\,\slash{S}_\sT\gamma_5\,[0;\lambda n]\,q(\lambda n)\,|P,S_\sT\rangle\,,\label{eq:gT}
\end{eqnarray}
where $P^\mu$ and $M$ are the nucleon's four momentum and mass, respectively, while $n^\mu$ is a light-cone vector with $n^2=0$ and $n\cdot P=1$, and $S_\sT$ refers to the transverse nucleon polarization vector. Furthermore, the definition (\ref{eq:gT}) contains quark fields $q$ and a Wilson line $[0;\lambda n]$ that ensures color gauge invariance. The twist-3 PDF $g_\sT^q$ can be expressed as a sum of a piece that is determined entirely in terms of twist-2 helicity PDF $g_1^q(x)$ (WW-contribution) and an interaction dependent {\it dynamical} twist-3 term, $\bar{g}_\sT^q(x)$, which involves quark-gluon correlations \cite{Wandzura:1977qf},
\begin{eqnarray}\label{decomposition}
g_\sT^q(x)=g_\sT^{q,WW}(x)+\bar{g}^q_\sT(x)\,,\\ \nonumber
g_\sT^{q,WW}(x)=\int_x^1{}\dfrac{dy}{y}g_1^q(y).
\end{eqnarray}
The explicit form of $\bar{g}^q_\sT(x)$ in terms of quark-gluon correlations can be found in Eq.~(46) of Ref.~\cite{Kanazawa:2015ajw}. For simplicity, the contribution of a quark mass term has been neglected in Eq.~({\ref{decomposition}). 

The $x^2$ moment of the dynamical twist-3 term $\bar{g}^q_\sT(x)$ is called $d^q_2/3$,
\begin{equation}
\int_{-1}^1\,\mathrm{d}x \,x^2\, \bar{g}_\sT^q(x)=\dfrac{d_2^q}{3},
\end{equation}
and can be related to the following local matrix element \cite{Shuryak:1981pi,Jaffe:1989xx}\footnote{Note that a ``-''  sign appears in Eq.(\ref{localmatrix}) which was missing in Ref.\cite{Burkardt:2008ps}.},
\begin{equation}\label{localmatrix}
d_2^q=-\dfrac{1}{2M(P^+)^2S^x}\langle P,S_\sT|\overline{q}(0)\gamma^+gG^{+y}(0)q(0)|P,S_\sT\rangle,
\end{equation}
where a special choice for the vectors $P^\mu=P^+(1,0,0,1)/\sqrt{2}$ and $n^\mu=(1,0,0,-1)/(\sqrt{2}P^+)$ was assumed.  Furthermore, $S^x$  is the nucleon polarization in the $\hat{x}$ direction and $G^{+y}$ is the gluon field strength tensor.

Some experimental information on the twist-3 parton distribution $g_\sT$ $\--$ and consequently on the second moment $d_2^q$ $\--$ may be obtained from the $g_2$ structure functions in polarized deep-inelastic lepton-nucleon (DIS) scattering.

As described in Ref.~\cite{Burkardt:2008ps}, the local matrix element, $\langle P,S_\sT|\overline{q}(0)\gamma^+gG^{+y}(0)q(0)|P,S_\sT\rangle$ appearing in Eq.~(\ref{localmatrix}), has a semi-classical interpretation as the average transverse color Lorentz force acting on the struck quark in a DIS experiment at the instant after it has been hit by the virtual photon, i.e.,
\begin{equation}\label{force}
F^{q,y}(0)\equiv-\dfrac{1}{\sqrt{2}P^+}\langle P,S_\sT|\overline{q}(0)\gamma^+gG^{+y}(0)q(0)|P,S_\sT \rangle.
\end{equation}
Comparing Eq.~(\ref{localmatrix}) and Eq.~({\ref{force}) suggests a connection between $d_2$ and this force. In particular,
\begin{equation}
F^{q,y}(0)=\sqrt{2}MP^+S^xd^q_2.
\end{equation}

The two-dimensional Fourier transform of the twist-2 GPDs lead to impact paramater space distributions. On the other hand the $x^2$ moments of the intrinsic twist-3 PDF $g_\sT$ can be related to the transverse color Lorentz force. The main purpose of this paper is to combine these two ideas and explore a physical interpretation for the Fourier transform of $x^2$ moments of intrinsic twist-3 GPDs as  \textit{the distribution of the average transverse color Lorentz force} on the transverse plane.

\section{ Color Lorentz Force Distribution in the Transverse Plane}
By taking the second moment of an intrinsic twist-3 PDF one can express this object in terms of a matrix element of a local operator that includes the covariant derivative $n\cdot D=n\cdot\partial -ign\cdot A=(\partial^+-igA^+)/P^+$ acting on quark fields. In particular, the antisymmetric combination is relevant, $\bar{q}\overleftrightarrow{D}^+ q=\frac{1}{2}(\bar{q}(D^+q)-(D^+\bar{q})q)$.
For example the $x^2$ moment of the chirally-even spin-dependent twist-3 parton distribution,  $g_{\sT}^q(x)$, can be written as,
\begin{equation}
\label{gT}
\int_{-1}^1 \mathrm{d}x \,x^2\, g_{\sT}^q(x)=
-\dfrac{1}{2S^xMP^{+2}}\langle P,S_\sT|\overline{q}(0)\gamma^x\gamma_5 (\overleftrightarrow{D}^{+})^2q(0)|P,S_\sT\rangle.
\end{equation}
If we compare this local matrix element with the second moment of the RHS of Eq.~(\ref{decomposition}), we find
\begin{equation}
\label{D-forward}
\langle P,S_\sT|\overline{q}(0)\gamma^x\gamma_5(\overleftrightarrow{ D}^+)^2q(0)|P,\lambda\rangle 
=-2S^xMP^{+2}\int_{-1}^1\mathrm{d}x\,x^2\,g_\sT^{q,WW}(x)+\dfrac{1}{3}\langle P,\lambda|\overline{q}(0)\gamma^+gG^{+y}(0)q(0)|P,\lambda\rangle.
\end{equation}

The main idea is to generalize the forward matrix elements in Eq.~(\ref{D-forward}) to non-forward matrix elements $\--$ which is possible if there is no momentum transfer in the $n$-direction from the initial state to the final state, or, in other words, the GPD skewness parameter $\xi$ vanishes, i.e. $\xi=0$. In this way, additional information on the position dependence of the transverse color Lorentz force can be obtained.
The non-forward generalization of Eq.~(\ref{D-forward}) has the following form,
\begin{equation}
\label{D-nonforward}
\langle p',\lambda'|\overline{q}(0)\gamma^x\gamma_5 (\overleftrightarrow{D}^+)^2q(0)|p,\lambda\rangle\Big|_{\xi=0}=\int_{-1}^{1} \mathrm{d}x\, x^2\, \mathrm{GPD^{WW}}(x,0,-{\bf \Delta}_\perp^2)+\dfrac{1}{3}\langle p',\lambda'|\overline{q}(0)\gamma^+gG^{+y}(0)q(0)|p,\lambda\rangle|_{\xi=0}.
\end{equation}
In analogy to Eq.~(\ref{decomposition}) we expect the Wandzura-Wilczek (WW) term in (\ref{D-nonforward}) to consist of twist-2 GPDs only. The second term on the RHS of (\ref{D-nonforward}) is the non-forward matrix element of the same operator that provides the average force in Eq.~(\ref{force}). This suggests that the transverse force distribution can be studied by the  $x^2$ moments of  twist-3 GPDs.
 Using transversely localized states $\left|P^+,{\bf R}_\perp =0,\lambda \right\rangle \equiv {\cal N} \int d^2p_\perp
 \left|P^+,{\bf p}_\perp,\lambda \right\rangle $, where ${\cal N}$ is a normalization factor, 
 the transverse force distribution can be defined as,
\begin{eqnarray}
\label{eq:Fi}
\mathcal{F}^i_{\lambda^\prime \lambda}({\bf b}_\perp)& \equiv& -\dfrac{1}{\sqrt2 P^+}\left\langle P^+,{\bf R}_\perp =0,\lambda^\prime \right|\overline{q}({\bf b}_\perp)\gamma^+gG^{+i}({\bf b}_\perp)q({\bf b}_\perp)\left|P^+,{\bf R}_\perp =0,\lambda \right\rangle
\end{eqnarray}
Therefore, just as the Fourier transform of the twist-2 GPDs gives the longitudinal momentum distribution in the transverse plane in Eq.~(\ref{FTtwist2}), the Fourier transform of the non-forward local matrix element (\ref{eq:Fi}) gives the distribution of the force in the transverse plane, i.e.,
\begin{equation}\label{Force}
\mathcal{F}^i_{\lambda'\lambda}({\bf b}_{\perp})=\int{}\dfrac{d^2 {\bf \Delta}_{\perp}}{(2\pi)^2}e ^{-i{\bf b}_{\perp}\cdot {\bf \Delta}_{\perp}} F^i_{\lambda'\lambda}({\bf \Delta}_{\perp})
\end{equation}
with,
\begin{equation}
\label{G-nonforward}
F^i_{\lambda'\lambda}({\bf \Delta}_{\perp})=-\dfrac{1}{\sqrt{2}P^+}\langle P^+,\tfrac{{\bf \Delta}_{\perp}}{2},\lambda'|\overline{q}(0)\gamma^+gG^{+i}(0)q(0)|P^+,-\tfrac{{\bf \Delta}_{\perp}}{2},\lambda\rangle.
\end{equation}

\section{Form Factors of $\overline{q}Gq$ Correlator}

In order to gain further insight into the physical meaning of the impact parameter Lorentz-force matrix element (\ref{Force}) we parameterize the general matrix element, 
\begin{eqnarray}
W^{\mu,\nu\lambda}(p,p^\prime)=\langle p',\lambda'|\bar{q}(0)\gamma^{\mu}igG^{\nu\lambda}(0) q(0)|p,\lambda\rangle\,,\label{eq:W}
\end{eqnarray}
 in terms of form factors. The way to parameterize $W$ closely follows the procedure outlined in Ref.~\cite{Meissner:2009ww}. In short, we assume the following ansatz for the matrix element $W$,
\begin{equation}
 W^{\mu,\nu\lambda}=\bar{u}(p',\lambda')\Gamma^{\mu ,\nu\lambda}(p,p')u(p,\lambda),
\end{equation}
where $\Gamma$ is a general Dirac matrix depending on the initial and final nucleon momenta, $p$ and $p'$, respectively. After  decomposing  $\Gamma$  into the sixteen basis matrices, $\{1,\gamma_5, \gamma^{\mu}, \gamma^{\mu}\gamma_5,i\sigma^{\mu\nu}\}$,  the coefficients in that decomposition are parameterized in terms of the four momenta $p^\mu$ and $p^{\prime \mu}$, along with form factors that depend on $t=\Delta^2=(p^\prime-p)^2$. A parameterization with a minimal number of form factors is obtained by applying parity, time reversal and hermiticity constraints, as well as Gordon identities.

For general Lorentz indices the matrix elements of the operator $\bar{q}(0)\gamma^{\mu}igG^{\nu\lambda}(0)q(0)$ can be parameterized in terms of 8 form factors \cite{in prep}. However, for the transverse force distribution, we are only interested in the matrix elements of $\bar{q}(0)\gamma^+igG^{+i}(0)q(0)$ which can be parameterized in terms of 5 form factors, $\Phi_1(t),...,\Phi_5(t)$ as, 
  \begin{eqnarray}\label{FormFactors}
\langle p',\lambda'|\bar{q}(0)\gamma^{+}igG^{+i}(0) q(0)|p,\lambda\rangle
&=&\overline{u}(p',\lambda')\Big\{\dfrac{1}{M^2}[P^{+}\Delta_{\perp}^i-P^{\perp}\Delta^{+})]\gamma^+\Phi_1(t)+\dfrac{P^+}{M}i\sigma^{+i}\Phi_2(t)\\ \nonumber
&+&\dfrac{1}{M^3} i \sigma ^{+\Delta}\big[P^+\Delta_{\perp}^i\Phi_3(t)-P^\perp \Delta^+\Phi_4(t)\big]
+\dfrac{P^+\Delta^+}{M^3}i \sigma ^{i\Delta}\Phi_5(t) \Big\}u(p',\lambda)
\end{eqnarray}
Here, $i$ corresponds to a transverse index $i=x,y$.

Combining Eq.~(\ref{Force}) with Eq.~(\ref{FormFactors})  gives the spatial distributions of the force fields described by each form factor,
\begin{eqnarray}
\label{Fb}
\mathcal{F}^i_{\lambda^\prime \lambda}({\bf b}_\perp) &=& \dfrac{i}{\sqrt2 P^+}\int {}\dfrac{\mathrm{d}^2{\bf \Delta}_\perp}{(2\pi)^2}\,\mathrm{e}^{-i{\bf b}_\perp \cdot {\bf \Delta}_\perp} \bar{u}(p',\lambda^\prime)\Big[
\dfrac{P^+\Delta^i_\perp}{M^2}\gamma^+\,\Phi_1(-{\bf \Delta}_\perp^2) +\dfrac{P^+}{M}i \sigma^{+i}\,\Phi_2(-{\bf \Delta}_\perp^2) \\ \nonumber
&&+\dfrac{P^+\Delta_\perp^i}{M^3}i\sigma^{+\Delta}\,\Phi_3(-{\bf \Delta}_\perp^2)\Big] u(p,\lambda).
\end{eqnarray}
As for $q(x,b_{\perp})$ in Eq.~(\ref{FTtwist2}), the limit $\xi\rightarrow0$  is necessary to develop a position space interpretation. Therefore, the $\Delta^2$ dependence of the form factors reduces to a ${\bf \Delta}_{\perp}^2$ dependence in Eq.~(\ref{Fb}). Since $\Delta^+=0$ for $\xi=0$, the form factors $\Phi_4$, $\Phi_5$ and the second term in the coefficient of  $\Phi_1$ in Eq.~(\ref{FormFactors}) do not appear in Eq.~(\ref{Fb}). 

The force fields resulting from a Gaussian toy model for the form factors are depicted in Fig.~\ref{fig:Fs}.
\begin{figure}[ht]
\includegraphics[width=7.5cm]{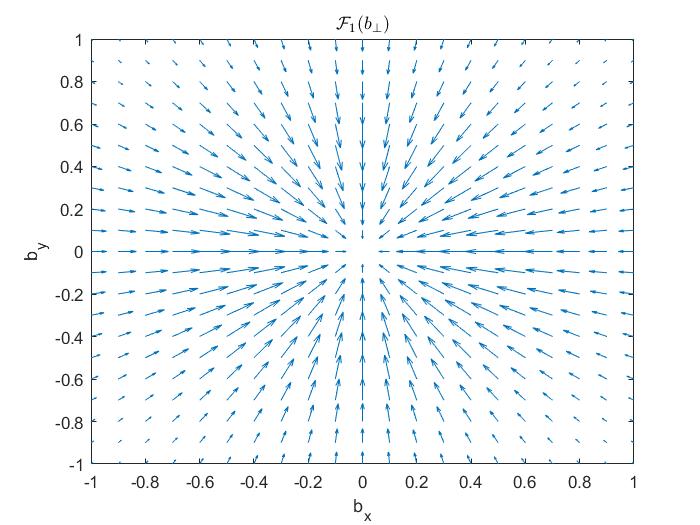}
\hspace{0.1cm}
\includegraphics[width=7.5cm]{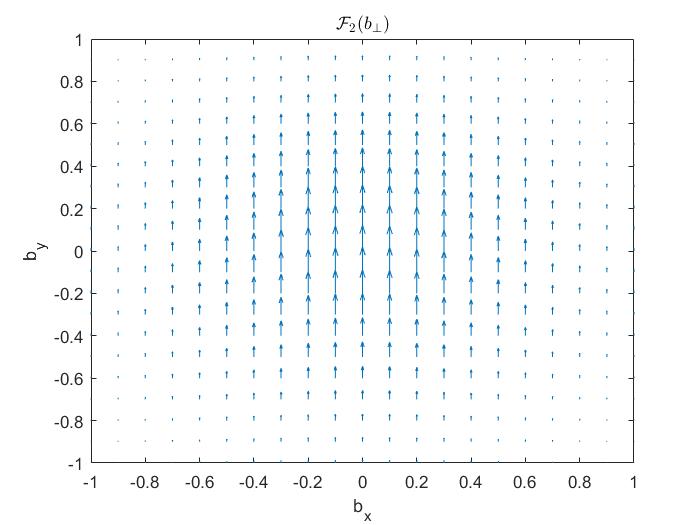}
\hspace{0.1cm}
\includegraphics[width=7.5cm]{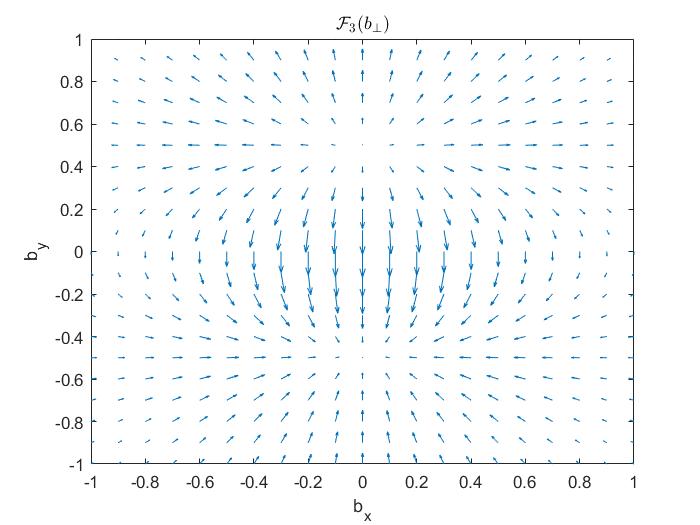}
\caption{ Transverse force fields obtained from the three form factors, $\Phi_1, \Phi_2$ and $\Phi_3$ in Eq.~(\ref{FormFactors}). The orientations of the force fields $\mathcal{F}_2$ and $\mathcal{F}_3$   are for a nucleon polarized in the $x$-direction.}   
\label{fig:Fs}
\end{figure}

Since $\bar{q}(0)\gamma^+gG^{+i}(0)q(0)$ is not sensitive to the polarization of quarks, the form factors describe forces on unpolarized quarks. Our comments on each term in Eq.~(\ref{Fb}) are as follows,

\begin{itemize}
  \item The first term involving $\bar{u}(p^\prime,\lambda^\prime)\,\gamma^+\,u(p,\lambda)$ in Eq.~(\ref{Fb}) is diagonal in the helicities $\lambda,\lambda^\prime$ and therefore not sensitive to the nucleon polarization. Thus the Fourier transform of  $\Delta_\perp^i\, \Phi_1$ yields the distribution of the axially symmetric {\it radial} force acting on  unpolarized quarks in an unpolarized nucleon ($\mathcal F_1({\bf b}_\perp$)).

\item  The second term in (\ref{Fb}) involving $\bar{u}(p^\prime,\lambda^\prime)\,i\sigma^{+i}\,u(p,\lambda)$ requires a nucleon helicity flip and it is thus sensitive to the transverse polarization of the nucleon. Therefore a Fourier transform of $\Phi_2$ describes the transverse force acting on unpolarized quarks in a transversely polarized nucleon and leads to the spatial distribution of the Sivers force ($\mathcal F_2({\bf b}_\perp$)).

 \item The third term in (\ref{Fb}) involving $\bar{u}(p^\prime,\lambda^\prime)\,i\sigma^{+\Delta}\,u(p,\lambda)$ also requires a nucleon helicity flip and  depends on the transverse nucleon polarization as well. The position dependence described by a Fourier transform of $\Phi_3$ is similar to the transverse Lorentz force $\vec{v}\times \vec{B}$ for a charged particle moving through a magnetic dipole field ($\mathcal F_3({\bf b}_\perp$)).

\end{itemize}
\section{Summary and Discussion}

Taking $x^2$ moments of twist-3 PDFs provides information about forward matrix elements of local quark-gluon-quark correlators that have
a very intuitive interpretation as the average transverse force acting on the active quark in a DIS experiment after absorbing the virtual photon. Similarly, $x^2$ moments of twist-3 GPDs  yield non-forward matrix elements of the same local quark-gluon-quark correlator that appears in $x^2$ moments of twist-3 PDFs.

We have shown that by taking a Fourier transform of these non-forward matrix elements, one can determine how the transverse force
depends on the impact parameter, $\bf{b}_\perp$.

Even though twist-3 GPDs are  difficult to extract from experiment, the relevant matrix elements can also be obtained from lattice QCD calculations. The related form factors in Eq.~(\ref{FormFactors}) can be extracted by considering the non-forward matrix elements of the same operator that is used to calculate $d_2$ \cite{Gockeler:2005vw}.

{\bf Acknowledgements:}
This work was partially supported by the DOE under grant number 
DE-FG03-95ER40965 (F. Aslan and M. Burkardt), and within the framework of the TMD Topical Collaboration.


\begin{thebibliography}{99}

\bibitem{Burkardt:2000za} 
  M.~Burkardt,
  Phys.\ Rev.\ D {\bf 62}, 071503 (2000)
  Erratum: [Phys.\ Rev.\ D {\bf 66}, 119903 (2002)]
  doi:10.1103/PhysRevD.62.071503, 10.1103/PhysRevD.66.119903
  [hep-ph/0005108].
  
  
\bibitem{Burkardt:2008ps} 
  M.~Burkardt,
  Phys.\ Rev.\ D {\bf 88}, 114502 (2013)
  doi:10.1103/PhysRevD.88.114502
  [arXiv:0810.3589 [hep-ph]].

\bibitem{Kanazawa:2015ajw}
 K.~Kanazawa, Y.~Koike, A.~Metz, D.~Pitonyak and M.~Schlegel,
 Phys.\ Rev.\ D {\bf 93}, 054024 (2016),
 doi:10.1103/PhysRevD.93.054024,
[arXiv:1512.07233 [hep-ph]].



\bibitem{Wandzura:1977qf} 
  S.~Wandzura and F.~Wilczek,
  Phys.\ Lett.\  {\bf 72B}, 195 (1977).
  doi:10.1016/0370-2693(77)90700-6
  
  
\bibitem{Shuryak:1981pi} 
  E.~V.~Shuryak and A.~I.~Vainshtein,
  Nucl.\ Phys.\ B {\bf 201}, 141 (1982).
  doi:10.1016/0550-3213(82)90377-7
  
  
\bibitem{Jaffe:1989xx} 
  R.~L.~Jaffe,
  Comments Nucl.\ Part.\ Phys.\  {\bf 19}, no. 5, 239 (1990).

\bibitem{Meissner:2009ww}
     S.~Meissner, A.~Metz and M.~Schlegel,
    JHEP {\bf 08}, 056 (2009), 
    doi:10.1088/1126-6708/2009/08/056,
    [arXiv:0906.5323 [hep-ph]].

  
  
\bibitem{in prep}
Aslan, Burkardt, Schlegel, in preparation.


\bibitem{Gockeler:2005vw} 
  M.~Gockeler, R.~Horsley, D.~Pleiter, P.~E.~L.~Rakow, A.~Schafer, G.~Schierholz, H.~Stuben and J.~M.~Zanotti,
  Phys.\ Rev.\ D {\bf 72}, 054507 (2005)
  doi:10.1103/PhysRevD.72.054507
  [hep-lat/0506017].

\end{thebibliography}
\end{document}